\documentclass[conference]{IEEEtran}
\makeatletter
\def\ps@headings{%
\def\@oddhead{\mbox{}\scriptsize\rightmark \hfil \thepage}%
\def\@evenhead{\scriptsize\thepage \hfil \leftmark\mbox{}}%
\def\@oddfoot{}%
\def\@evenfoot{}}
\makeatother
\usepackage[utf8]{inputenc}
\usepackage{amsmath,amssymb,amsfonts}
\usepackage{algorithmic}
\usepackage{graphicx}
\usepackage{textcomp}
\usepackage{booktabs}
\usepackage{multirow}
\usepackage{multicol}
\usepackage{pifont}
\usepackage{subfigure}
\usepackage{graphicx}
\usepackage{comment}
\usepackage{balance}
\usepackage{amsmath}
\usepackage{color}
\usepackage{url}
\usepackage[linesnumbered,ruled,vlined]{algorithm2e}
\SetKw{KwBy}{by}
\SetKw{KwTo}{in}
\usepackage{mathtools}

\usepackage[table,xcdraw]{xcolor}
\usepackage{ulem}
\usepackage{cancel}
\usepackage[hyperindex,breaklinks]{hyperref}
\usepackage{subfigure}

\newcolumntype{P}[1]{>{\centering\arraybackslash}p{#1}}

\newcommand*{\affmark}[1][*]{\textsuperscript{#1}}

\begin{document}

\title{An Incremental Gray-box Physical Adversarial Attack on Neural Network Training}

\author{\IEEEauthorblockN{Rabiah Al-qudah\affmark[1], Moayad Aloqaily\affmark[1], Bassem Ouni\affmark[2], Mohsen Guizani\affmark[1], Thierry Lestable\affmark[2]} \\

\IEEEauthorblockA{
\affmark[1]Mohamed Bin Zayed University of Artificial Intelligence (MBZUAI), UAE \\
\affmark[2]Technology Innovation Institute, Abu Dhabi, UAE \\
E-mails: \protect \affmark[1]\{rabiah.alqudah; moayad.aloqaily; mohsen.guizani\}@mbzuai.ac.ae}
\affmark[2]\{bassem.ouni; thierry.lestable\}@tii.ae\\

\vspace{-1.0cm}

}

\maketitle

\begin{abstract}
Neural networks have demonstrated remarkable success in learning and solving complex tasks in a variety of fields. Nevertheless, the rise of those networks in modern computing has been accompanied by concerns regarding their vulnerability to adversarial attacks. In this work, we propose a novel gradient-free, gray box, incremental attack that targets the training process of neural networks. The proposed attack, which implicitly poisons the intermediate data structures that retain the training instances between training epochs acquires its high-risk property from attacking data structures that are typically unobserved by professionals. Hence, the attack goes unnoticed despite the damage it can cause. Moreover, the attack can be executed without the attackers' knowledge of the neural network structure or training data making it more dangerous. The attack was tested under a sensitive application of secure cognitive cities, namely, biometric authentication. The conducted experiments showed that the proposed attack is effective and stealthy. Finally, the attack effectiveness property was concluded from the fact that it was able to flip the sign of the loss gradient in the conducted experiments to become positive, which indicated noisy and unstable training. Moreover, the attack was able to decrease the inference probability in the poisoned networks compared to their unpoisoned counterparts by 15.37\%, 14.68\%, and 24.88\% for the Densenet, VGG, and Xception, respectively. Finally, the attack retained its stealthiness despite its high effectiveness. This was demonstrated by the fact that the attack did not cause
a notable increase in the training time, in addition, the Fscore values only dropped by an average of 1.2\%, 1.9\%, and 1.5\% for the poisoned Densenet, VGG, and Xception, respectively.
\end{abstract}


\section{Introduction}\label{sec:intro}
Cognitive cities \cite{cogn1} are proactive, hyper-connected, and citizen-driven cities that are designed to minimize resource consumption, in order to achieve sustainability. In addition, the vast advancement in Artificial Intelligence (AI) and Internet of Things (IoT) technologies have enhanced the evolution of research that integrates both technologies to deliver and automate services for cognitive cities' residents. In fact, the great development that emerged from the integration of those technologies has brought unforeseen exposures to cybersecurity, in addition to novel attacks that need to be addressed in order to deliver secure automation to cognitive cities.

Securing access to different services and facilities, such as connected buildings and data centers, and managing the flow of foot traffic are crucial requirements when adopting the cognitive city paradigm. Those requirements can be implemented using biometric authentication such as fingerprint recognition and iris recognition. Despite the benefits of biometric authentication, privacy concerns and security attacks pose serious challenges to this technology after deployment. Attacks that target biometric recognition systems typically include presenting human characteristics or artifacts directly to a biometric system to interfere or bias with its standard operation. Such attacks can result in granting access to unauthorized individuals into secured premises, allowing tailgating, or triggering denial of service by rejecting the biometrics of authorized individuals. For instance, in 2017, the Chaos Computer Club executed a successful attack on the Samsung Galaxy S8 iris scanner using a simple photograph and a contact lens \cite{samsung}.

On a different note, neural networks have gained wide popularity in the past decade due to their supremacy in terms of accuracy and minimal need for human intervention. Moreover, those networks are data hungry and are very sensitive to patterns they are exposed to during the training phase. On the other hand, neural networks are vulnerable and can be biased even with the introduction of simple adversarial attacks. For example, altering a single pixel in the data fed to an image classifier can disrupt the learning experience and result in a biased model \cite{lit2}.

Adversarial attacks are considered white box when the attacker has full access to the neural network and data, while gray box attacks assume having access to either and black box attacks assume access to neither. Those attacks can be categorized into digital
attacks and physical attacks. Digital attacks engineer pixel values of input images, whereas physical attacks insert pixel patches that represent real world objects into the input image instance. Attacker goals can vary from faulting the predictions of a certain class, in what is called ``targeted attacks''. Moreover, an attack can be ``non-targeted'' and aim to fault the model in general. 

Furthermore, attacks that target faulting the inference phase have been extensively studied in the literature. On the contrary, only a handful of papers focused on faulting the training phase and the intermediate values related to its computations. In 2022, Breier \textit{et al.} introduced the first attack that directly targets the training phase by perturbing the ReLu values while training  \cite{whiletraining}. In fact, the lack of research attention on attacks that target the training phase puts many applications that rely on neural networks in jeopardy. In this work, we propose and test a novel attack that focuses on faulting the training process of neural networks in the domain of biometric authentication through iris recognition. The contributions of this work can be summarized as follows:
\begin{enumerate}
\item We introduce a novel gradient-free, data poisoning attack that incrementally and directly targets the training set during the training process of a neural network with minimal knowledge by the attacker. To the best of our knowledge, this is the first attack that executes between training epochs and targets the intermediate data structures of the training phase.
\item We conduct extensive experimental verification on the proposed attack to test its effectiveness and stealthiness. We define four evaluation criteria to quantify the effect of the attack, namely, the average of the loss change, the average inference probability, the training time difference, and the performance degradation measure.
\item We experiment the proposed attack on an important aspect of a cognitive city, namely, iris recognition. To the best of our knowledge, this is the first attempt to test the effect of an adversarial attack that occurs during training on the domain of iris recognition. 
\end{enumerate}
The rest of this paper is organized as follows: the most recent literature on the domain of physical attacks and iris recognition is presented in Section \ref{sec:related work}. The proposed methods are outlined in Section \ref{method}. The results are described and discussed in Section \ref{sec:results}. Finally, Section \ref{sec:conclusion} concludes and summarizes the main highlights and observations of this work.
\section{Related Work}\label{sec:related work}
\subsection{Attacks on Neural Networks}
Patch attacks are physical attacks that replace a subset of pixels in an image with pixels from adversarial patches to bias a model \cite{ lit3}. While many studies have proposed attacks that target faulting the inference phase \cite{litinf1,litinf2}, only a handful of papers focused on faulting the training phase and the intermediate values related to its computations \cite{whiletraining}. For example, Zhao \textit{et al.} \cite{litinf1} applied the alternating direction method of multipliers at the inference time to solve the optimization problem of the targeted fault sneaking attack. The results showed that the attack was successful and stealthy, moreover, the success rate was approximately 100\% when the number of targeted images was less than 10. Whereas, the success rate decreased as the number of fooled images increased. Furthermore, the work in \cite{litinf2} studied the effects of bitwise perturbations at inference time on 19 deep networks. The vulnerable parameters of the experimented networks were identified using heuristic functions. The results showed that most deep architectures have at least one parameter that causes an accuracy loss of over 90\% when a bit-flip is executed on their bitwise representation. 

In addition, the Fast Gradient Sign
Method (FGSM) has been widely used in the literature as an attacking strategy \cite{FGSM}. This method includes adding noise whose direction is the same as the gradient of the cost function with respect to the data using a trained model. The work in \cite{whiletraining}, proposed the first attack that targets the training phase by changing the values of the ReLu function to bias the neural network. The novel attack was proven to be effective and stealthy.
\subsection{Attacks on Iris Recognition Systems}
The crucial role iris recognition has played in securing premises, in addition to the threatening effects of breaching such authentication systems, have made iris biometric authentication systems an active target for adversarial attacks. A novel morph attack on iris recognition systems was tackled in \cite{lit7spoof}. Sharma \textit{et al.} generated morphed iris images using the Indian Institute of Technology Delhi (IITD) Iris Database and West Virginia University (WVU) multi-modal datasets. The morph attack achieved a success rate higher than 90\% on two state-of-the-art iris recognition methods, which indicates the vulnerability of iris recognition systems.

In order to protect against the increasing attacks, researchers have also focused on studying countermeasures and detection mechanisms for iris recognition attacks. For example, Thukral \textit{et al.} \cite{lit6spoof} proposed an iris spoofing detection system that utilized Gabor filters and Histogram of Gradient (HOG) bins to extract features. Next, a Support Vector
Machine (SVM) was used to detect if the extracted features represented fake or real iris. The proposed system was able to detect spoofing attacks with an accuracy of 98\%. Finally, Tapia \textit{et al.} \cite{lit5spoof} tackled testing the liveness of the scanned iris to protect the system from being fooled by printed images or artificial eyes. The proposed work utilized a MobileNetV2 network, which was trained from scratch. Moreover, the authors increased the number of filters and weighted each class based on the number of its instances. The proposed method was able to accurately classify irises with competitive Bona Fide Classification Error Rates (BPCER) of less than 4\% in all experiments.
\section{Physical Gray-box Adversarial Attacks}\label{method}
A labeled training set of size $s$ can be represented as $DS =\{(x^i$, $y^i)\}^{s}_{i=1}$, where $y^i \in \mathcal{Y}$ and $\mathcal{Y}$ is the set of all possible output classes for an image classification problem. When training a deep classifier, we aim to optimize a discriminant function $\mathcal{F}$ that maps each instance,$x^i$, to the class associated with the highest class probability, as can be seen in Equation \ref{eq:softmaxeq}. This optimization process takes place during the training process by passing $DS$ to a deep classifier for a number of training rounds. The number of training rounds will be referred to as $Epochs$ throughout the rest of this paper. The aforementioned setting of training $\mathcal{F}$ without any attacks will be referred to as the \textbf{base model} throughout this work.
\begin{equation} \label{eq:softmaxeq}
\mathcal{F} \rightarrow argmax(P(\mathcal{Y} \mid x^i))
\end{equation}
\subsection{Attack Definition}\label{attackdef}
In our proposed attack, an attacker aims to corrupt the training process by perturbing the training instances incrementally between training epochs in order to optimize a corrupted poisoned discriminant function $\mathcal{F'}$ that produces faulty probability distributions over the possible output classes. The attack is executed implicitly in multiple rounds. In each poisoning round, a poisoning procedure that selects $X \subseteq DS$ of size $|X|= \alpha*s$ is executed, where $\alpha \in (0\%,100\%]$ is the poisoning percentage coefficient. The attacker's goal is to replace $X=\{(x^i$, $y^i)\}^{|X|}_{i=1}$ with a poisoned set $X'=\{g(x^i$), $y^i)\}^{|X|}_{i=1}$, where g(.) is the poisoning function that modifies $x^i$ at a pixel level. The poisoning function replaces the pixels that fall within a selected area, namely $Patch_{Area}$, with faulty pixels, $x'$, in order to corrupt the image representation and result in a faulty training process. The poisoning function can be seen in Equation \ref{eq:poisoningfun}, where $W$ and $H$ are the width and height of the training image instance $x^i$.
\begin{equation} \label{eq:poisoningfun}
g(x)=
\begin{cases} 
      x'_{u,v}   & if (u,v) \in Patch_{Area}\\
      & \hspace{3 cm} ,u \in [0,W) , v \in [0,H)\\
      x_{u,v} & Else
\end{cases}
\end{equation}
The attack targets the intermediate data structures, where the training instances are saved. In addition, it is executed incrementally between training epochs, such that a different $X$ is selected every poisoning round in order to accumulate the poisoned training instances and increase the effectiveness of the attack. 

The attack frequency coefficient determines the number of poisoning rounds and is called $\beta \in [1,Epochs]$. When the value of $\beta$ is chosen to be 1, then the attack will be executed after each training epoch causing an increased risk of damage. On the contrary, if the value is chosen to be $Epochs$, then the poisoning process will only happen once after the first training epoch.
\subsection{Poisoning Strategy}\label{strategysec}
Function $g(.)$ in Equation \ref{eq:poisoningfun} replaces pixels in a training instance within the defined poisoning area, $Patch_{Area}$. This poisoning procedure can be implemented in multiple ways. In this work, we opted to implement $g(.)$ to execute local perturbations and global perturbations \cite{localglobalp}. It is worth mentioning that only one type of perturbations was considered in each of the conducted experiments in this work.

In the local perturbations setting, a small area called physical patch in the training instance is selected and replaced with pixels from another image. In this work, the physical patch was chosen to be close to the training set domain, hence it was an image of a human eye. It is worth mentioning that the size of the $Patch_{Area}$ and its location are randomized, and optimizing them is out of the scope of this work \cite{lit3}. 

On the other hand, in the global perturbations setting, all the instances in $X$ are replaced with another randomly selected image from the training set. This way the classifier will be exposed to a highly redundant training set which corrupts the training process by increasing the risk of overfitting. Both poisoning strategies are not easy to blacklist, since the local setting only alters a small area of each instance and the global perturbation setting uses an image from within the training instances in a manner that imitates image augmentation, which is a benign, widely used technique in training neural networks.
\subsection{Attack Characteristics}\label{attackspec}
The attack specifications can be summarized as:
\begin{enumerate}
    \item \textbf{The attack is non-targeted:} the attack definition in \ref{attackdef} shows that no restrictions apply on the choice of $y^i$ in $X$. Moreover, the value of $y^i$ remains unchanged after poisoning takes place in $X'$.
    \item \textbf{The attack does not affect packets delay:} the attack only targets the training phase, whereas the inference phase is executed in the usual manner. Hence, the attack is stealthy in the sense that it does not affect the packet delay when the deep classifier is deployed on the cloud.
     \item \textbf{The attack samples without replacement:} to guarantee faster and stealthier execution, $X$ is sampled every poisoning round without replacement; that is an instance can only be included once in $X$ at a certain poisoning round, however an instance can be included in multiple poisoning rounds. This implies that the network will be exposed to a different training set after every poisoning round, which results in a higher training instability. 
     \item \textbf{The attack is incremental for increased effectiveness:} the poisoned instances in $X'$ accumulate in the training set after each poisoning round and throughout the training phase, which in turn intensifies the effect of poisoning even at a low value of $\alpha$.
    \item \textbf{The attack is gradient-free \cite{gradient-free} and is gray box:} the attack is gray box since we assume that the attacker only has access to the intermediate data structures of the training process without the need to access the physical path of the training instances or the neural network architecture. In other words, the attack is agnostic to the neural network architecture. The attack is also gradient-free since it perturbs the training data between epochs without the need to access the gradients of the attacked neural network. 
    \item \textbf{The attack targets intermediate data structures:} typically developers' efforts are more focused on preparing and preprocessing the training set before training. On the other hand, what happens during training and the values of the intermediate data structures that keep the training instances are overlooked, especially that the training is usually
conducted on powerful servers with limited physical access. Hence, this attack which poisons the data implicitly between training epochs, acquires its high risk property from attacking data structures that are typically not monitored by professionals, and hence the attack goes unnoticed despite the damage it causes.
\end{enumerate}
\subsection{Evaluation Metrics}\label{metricssec}
In each experiment, the neural networks will be evaluated and compared in terms of the following evaluation measures:
\begin{enumerate}
    \item Attack effectiveness measures: an attack is called effective if it achieves its intended goals. In our proposed attack, the goal is to expose the deep classifier to an unstable training process, which in turn, will result in faulty probability distributions produced by the network at the inference stage.
    \begin{enumerate}
        \item Average of Loss Change $(ALC)$: the loss function is typically expected to decrease as the training process progresses. This is due to backpropagation, which reflects what the network learned during each training epoch. The $ALC$ measures the average change in the loss value over the training epochs, and the sign of this evaluation metric is a leading element, as it reflects whether the loss was decreasing or increasing throughout training. Executing the attack is expected to cause instability in the training process due to the noisy poisoned data and, hence, increase the $ALC$ value. The $ALC$ can be defined as follows, where the $\ell$ is the loss and $Epochs$ is the number of training epochs:
        \begin{equation} \label{eq:eq2}
            ALC =  \frac{\sum_{i=1}^{Epochs} (\ell_{i} - \ell_{i-1})}{Epochs-1}
         \end{equation}
        \item Average Inference Probability (AIP): the softmax function is typically used in the last layer of deep classifiers to normalize the output to a probability distribution over the possible output classes. Each test instance is classified as the class of the highest probability. In this evaluation criterion, we assess the effect of the attack on the probabilities produced by the model at the inference stage, as typically higher probabilities imply more confidence about the selected class. As a result, a decreased average probability reflects the effectiveness of the attack on the final output of the model. $AIP$ can be calculated using Equation \ref{eq:eqopti}, where $t^i$ is a test instance.
        \begin{equation} \label{eq:eqopti}
         AIP= Average(argmax(P(Y \mid t^i)))
        \end{equation}
    \end{enumerate}
    \item Attack stealthiness measures: an attack is called stealthy if the evaluation metrics of the corrupted classifier $\mathcal{F'}$ are close to the metrics of the base model $\mathcal{F}$ \cite{whiletraining}.
    \begin{enumerate}
        \item Training Time Difference $(TTD)$: training a neural network can be a lengthy process, especially when the training instances are large. Hence, it is crucial to ensure that executing the attack will not cause an observable added amount of time to the training phase, in order to keep the attack unnoticed. The $TTD$ measure can be defined as follows:
         \begin{equation} \label{eq:eq1}
            TTD = Training Time' - Training Time_{base}   
         \end{equation}
        where $Training Time_{base}$ is the time taken to train the base model, and $Training Time'$ is the training time when the neural network is trained with poisoned data.
        \item Performance Degradation Measure (PDM): in order to confirm the attack stealthiness, the metrics of the poisoned classifier need to be reasonably close to the metrics of the base classifier. In this evaluation criterion, the difference between the macro Fscore of the base model and each poisoned model is calculated, as described in Equation \ref{eq:eq3}, where $Fscore'$ is the Fscore of a poisoned model.
         \begin{equation} \label{eq:eq3}
            PDM =  {Fscore}_{base} - Fscore'
         \end{equation}
    \end{enumerate}
  
\end{enumerate}
\subsection{Datasets}\label{dataset}
The proposed attack perturbs images and hence can target any computer vision application. Nevertheless, we opted to apply it to an iris recognition dataset, due to the significance of this domain. The CASIA Iris Subject Ageing dataset \cite{IrisDatset} was considered in our experiments. This dataset was collected by the National Laboratory of Pattern Recognition (NLPR) in China in April 2009 and April 2013. In this work, the subset of CASIA Iris Subject Ageing which was collected in 2009 using the H100 sensor was chosen due to its high diversity and good size. The subset comprises 37912 instances of the left and right eyes of 48 individuals. The dataset instances pose some challenging scenarios, like glasses, partially closed eyes, Moreover, some instances have very low brightness. The cross-validation method was used to train and evaluate the neural networks, and 100 images from each user subject were randomly selected for the test dataset. 
\subsection{Technical and Experimental Setup}\label{TechSetup}
Three state-of-the-art deep classifiers, namely, Densenet, VGG, and Xception were considered for this work. Moreover, the number of epochs was set to 10, the cross entropy loss function was used and the networks were trained with a learning rate of .01 on the Google Colab Pro platform which utilizes NVIDIA GPUs. It is worth mentioning that the code of this work is available on Github \cite{Github}.

Each of the 3 considered deep classifiers were experimented with $\alpha$ values of 5\%, 10\%,15\%, and 20\%, as 20\% is typically the maximum poisoning percentage considered in the literature \cite{twentyperc}. In the experiments description and results, the local perturbations poisoning strategy will be referred to as $P$, and the global perturbations strategy will be referred to as $R$. 
\section{Results and Discussion}\label{sec:results}
\begin{figure*}[htb!]
\centering
{\includegraphics[width=1.0\linewidth, height=4.25cm]{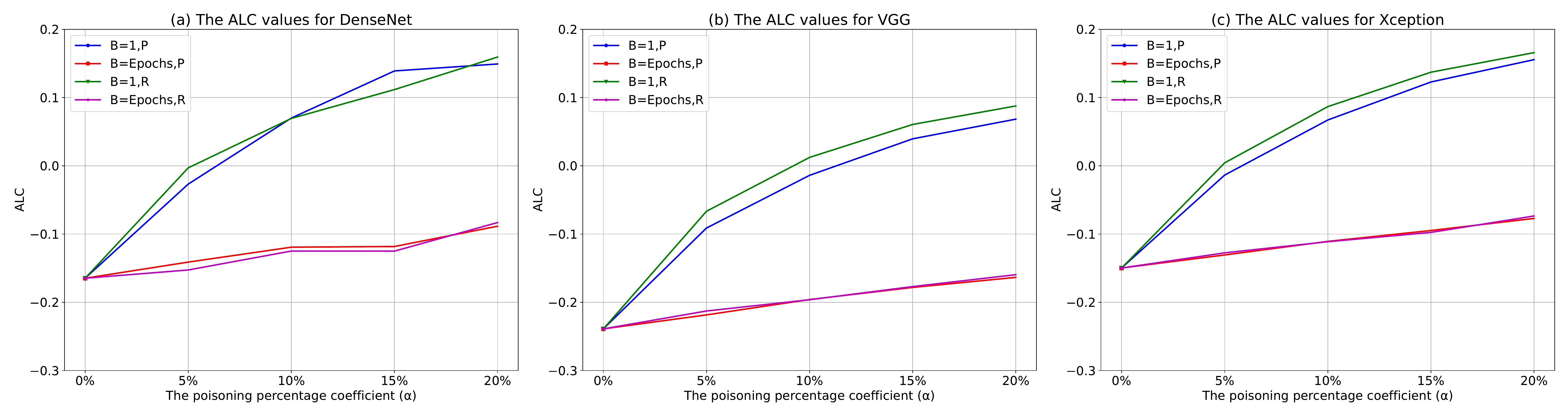}\label{fig:alc3}}
\caption{Experimental results of the Average of Loss Change (ALC) values
}
\label{fig:ALC}

\end{figure*}

\begin{figure*}[htb!]
\centering
{\includegraphics[width=1.0\linewidth, height=4.25cm]{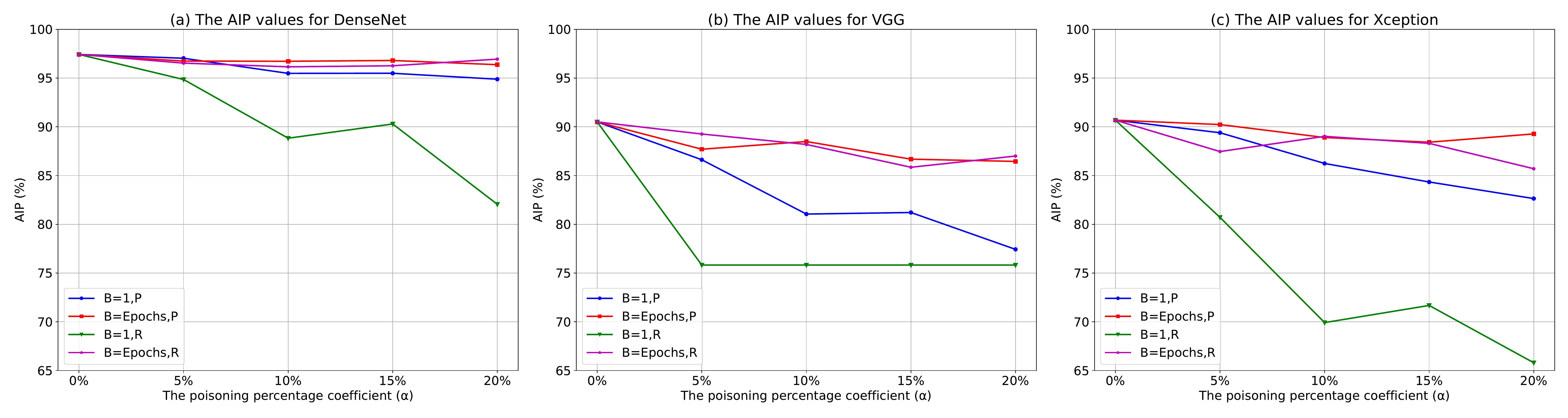}\label{fig:aip1}}
\caption{Experimental results of the Average Inference Probability (AIP) values}
\label{fig:AIP}

\end{figure*}

\begin{figure*}[htb!]
\centering
{\includegraphics[width=1.0\linewidth, height=4.25cm]{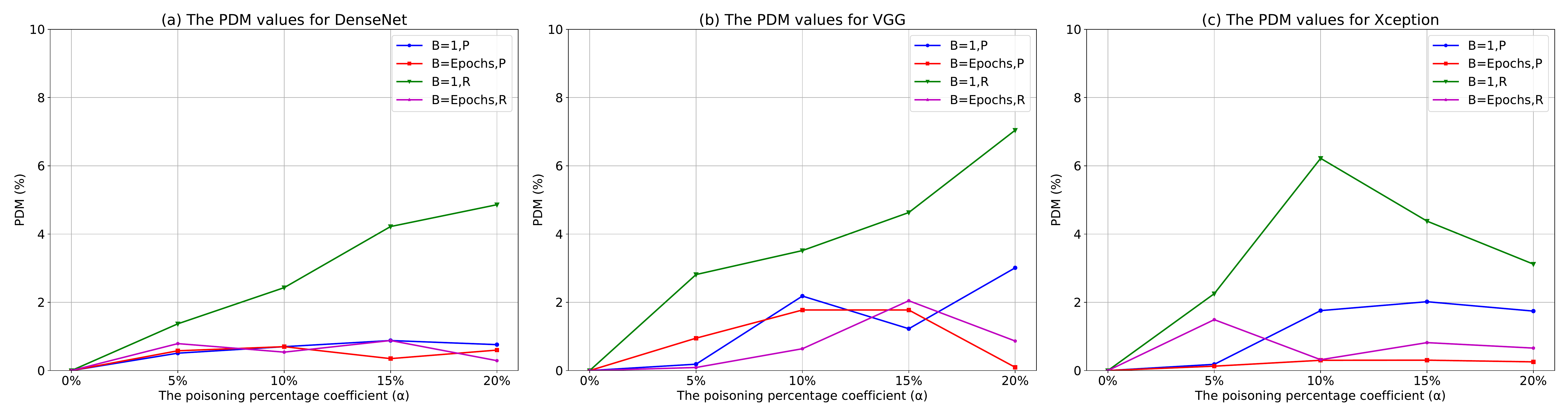}\label{fig:f1}}
\caption{Experimental results of the Performance Degradation Measure (PDM) values}
\label{fig:fscore}

\end{figure*}

In this section, the results of evaluating the proposed attack will be presented in detail. Figures \ref{fig:ALC}, \ref{fig:AIP}, and \ref{fig:fscore} depict the results of the evaluation metrics described in \ref{metricssec}. In all figures, the result of the base model is depicted as the origin point ($\alpha$ = 0\%).
\subsection{Analysis of Attack Effectiveness}
Figure \ref{fig:ALC} depicts the $ALC$ results for the 3 considered deep classifiers. A positive $ALC$ value indicates increasing loss values and poor training, whereas a low negative value indicates a more stable training process. From Figure \ref{fig:ALC}, it can be noted that increasing $\alpha$ was always associated with increased $ALC$ values, hence, it can be concluded that increasing the poisoning percentage increases the effectiveness of the attack.

On a different note, increasing the attack frequency (i.e., a lower $\beta$) resulted in increased effectiveness in all experiments. In the experiments where $\beta$'s value was set to 1, the $ALC$ kept increasing as the value of $\alpha$ increased, and the value was positive in all experiments where $\alpha \geq 10\%$. On the other hand, when $\beta = Epochs$, the $ALC$ results were increasing but negative in all experiments, which means that the loss values were still decreasing but at a lower rate compared to the base model and the experiments of higher frequency.

The $AIP$ results are depicted in Figure \ref{fig:AIP}, where it can be seen that increasing the value of $\alpha$ resulted in decreasing the $AIP$ in all experiments. However, this decrease varied in the experiments; for example, the decrease was slight, even when $\alpha$ increased, in the experiments where $\beta$=$Epochs$. On the other hand, increasing $\alpha$ with a higher frequency ($\beta = 1$) resulted in a more noticeable drop in the $AIP$ values. For example, it can be seen in Figure \ref{fig:AIP}(c) that the $AIP$ value dropped by 24.88\% when $\alpha=20\%$ and $\beta=1$ in the random poisoning experiment, $R$. Whereas, the $AIP$ value only dropped by 5\% when we only changed the value of $\beta$ to be equal to the number of $Epochs$. Furthermore, the highest drop in the $AIP$ in the poisoned networks
compared to their unpoisoned counterparts at inference time
was 15.37\%, 14.68\%, and 24.88\% for the Densenet, VGG,
and Xception, respectively. Overall, we can conclude that the attack was effective in all conducted experiments. Moreover, the attack effectiveness has a positive correlation with the percentage $\alpha$ and frequency $\beta$.
\subsection{Analysis of Attack Stealthiness}
It is crucial to keep the proposed attack undetected. The attack can be easily noticed if it takes long to execute, thus, to ensure the attack stealthiness, the $TTD$ measure is monitored in all experiments. Among all conducted experiments, the maximum $TTD$ value was 63 seconds. Hence, the attack did not add a noticeable period of time to the training time of the base model. Moreover, to monitor the stealthiness of the attack, the $PDM$ values were recorded as can be seen in Figure \ref{fig:fscore}. The maximum $PDM$ value was recorded for the VGG network with $\alpha =20\%$ and $\beta = 1$ in the random poisoning experiment, $R$. Overall, the average $PDM$ values were 1.2\%, 1.9\%, and 1.5\% for the Densenet, VGG, and Xception, respectively. Hence, it can be concluded that the attack demonstrated a stealthy behavior. 
\subsection{Analysis of Poisoning Strategy}
As explained in Section \ref{strategysec}, the attack was experimented under local perturbations setting ($P$) and global perturbations setting ($R$). The influence of the perturbation type was highly associated with the value of $\beta$. It can be seen in Figures \ref{fig:ALC}, \ref{fig:AIP} and \ref{fig:fscore} that in the experiments of low frequency, where $\beta = Epochs$, both perturbation types achieved comparable results. On the other hand, when the poisoning rounds were executed after every epoch, where $\beta$=1, the attack showed the highest effectiveness in the global perturbations setting, $P$.

Finally, the results showed that the proposed attack is effective and stealthy. Those properties increase when the attack is intensified by increasing the value of $\alpha$, increasing the number of affected pixels, similar to the case of global perturbations, and decreasing $\beta$ for higher execution frequency. Moreover, the proposed attack inherits its riskiness from attacking unobserved data structures that usually reside on powerful servers with limited physical access. The attack is also incremental and accumulates poisoned data gradually to intensify its effectiveness across the training epochs. In addition, the attack requires no knowledge about the neural network structure, as all experiments in this work were conducted using the same injection code.
\section{Conclusion and Future work}\label{sec:conclusion}
Neural networks are vulnerable to adversarial attacks. Moreover, the digital transformation adopted worldwide implies continuous acquisition and analytics of big streams of data, which has brought novel digital threats and unforeseen exposures to cybersecurity. In this work, we propose a novel gradient-free, gray box, incremental attack that targets the intermediate data structures of the training phase of neural networks. The attack has 3 main parameters: the attack percentage coefficient, the attack frequency coefficient, and the poisoning strategy. In all conducted experiments, it was noted that the attack stealthiness and effectiveness had a positive correlation with the aforementioned parameters.

Moreover, the attack resulted in unstable training, as it made the loss values increase which in turn indicates poor learning and generalization. Moreover, the attack was able to decrease the probability of the output class ($AIP$) in the poisoned networks compared to their unpoisoned counterparts at inference time by 15.37\%, 14.68\%, and 24.88\% for the Densenet, VGG, and Xception, respectively. Despite its effectiveness, the attack remained stealthy as it only dropped the Fscore values by 1.2\%, 1.9\%, and 1.5\% for the poisoned Densenet, VGG, and Xception, respectively.

In future works, further sensitivity analyses will be conducted on existing and new parameters, such as the type of communication protocol, and the area and size of the patch area. Moreover, the attack will be compared to other iris recognition attacks.
\section*{Acknowledgements}
This research was supported by the Technology Innovation Institute (TII), Abu Dhabi, UAE, under the CyberAI project (grant number: TII/DSRC/2022/3036).

\bibliographystyle{IEEEtran}
\bibliography{Ref}

\begin{thebibliography}{10}
\providecommand{\url}[1]{#1}
\csname url@samestyle\endcsname
\providecommand{\newblock}{\relax}
\providecommand{\bibinfo}[2]{#2}
\providecommand{\BIBentrySTDinterwordspacing}{\spaceskip=0pt\relax}
\providecommand{\BIBentryALTinterwordstretchfactor}{4}
\providecommand{\BIBentryALTinterwordspacing}{\spaceskip=\fontdimen2\font plus
\BIBentryALTinterwordstretchfactor\fontdimen3\font minus
  \fontdimen4\font\relax}
\providecommand{\BIBforeignlanguage}[2]{{%
\expandafter\ifx\csname l@#1\endcsname\relax
\typeout{** WARNING: IEEEtran.bst: No hyphenation pattern has been}%
\typeout{** loaded for the language `#1'. Using the pattern for}%
\typeout{** the default language instead.}%
\else
\language=\csname l@#1\endcsname
\fi
#2}}
\providecommand{\BIBdecl}{\relax}
\BIBdecl

\bibitem{cogn1}
J.~Machin, E.~Batista, A.~Ballesté, and A.~Solanas, ``Privacy and security in
  cognitive cities: A systematic review,'' \emph{Applied Sciences}, vol.~11, p.
  4471, 05 2021.

\bibitem{samsung}
A.~Morales, J.~Fierrez, J.~Galbally, and M.~Gomez-Barrero, ``Introduction to
  iris presentation attack detection,'' in \emph{Handbook of Biometric
  Anti-Spoofing, 2nd Ed.}, 2019.

\bibitem{lit2}
J.~Su, D.~V. Vargas, and K.~Sakurai, ``One pixel attack for fooling deep neural
  networks,'' \emph{IEEE Transactions on Evolutionary Computation}, vol.~23,
  no.~5, pp. 828--841, 2019.

\bibitem{whiletraining}
J.~Breier, X.~Hou, M.~Ochoa, and J.~Solano, ``Foobar: Fault fooling backdoor
  attack on neural network training,'' \emph{IEEE Transactions on Dependable
  and Secure Computing}, pp. 1--1, 2022.

\bibitem{lit3}
X.~Li and S.~Ji, ``Generative dynamic patch attack,'' 2021.

\bibitem{litinf1}
P.~Zhao, S.~Wang, C.~Gongye, Y.~Wang, Y.~Fei, and X.~Lin, ``Fault sneaking
  attack: a stealthy framework for misleading deep neural networks,''
  \emph{2019 56th ACM/IEEE Design Automation Conference (DAC)}, pp. 1--6, 2019.

\bibitem{litinf2}
S.~Hong, P.~Frigo, Y.~Kaya, C.~Giuffrida, and T.~Dumitra\c{s}, ``Terminal brain
  damage: Exposing the graceless degradation in deep neural networks under
  hardware fault attacks,'' in \emph{Proceedings of the 28th USENIX Conference
  on Security Symposium}, ser. SEC'19.\hskip 1em plus 0.5em minus 0.4em\relax
  USA: USENIX Association, 2019, p. 497–514.

\bibitem{FGSM}
V.~W. Anelli, A.~Bellogín, Y.~Deldjoo, T.~Di~Noia, and F.~A. Merra, ``Msap:
  Multi-step adversarial perturbations on recommender systems embeddings,''
  \emph{The International FLAIRS Conference Proceedings}, vol.~34, April 2021.

\bibitem{lit7spoof}
R.~Sharma and A.~Ross, ``Image-level iris morph attack,'' in \emph{2021 IEEE
  International Conference on Image Processing (ICIP)}, 2021, pp. 3013--3017.

\bibitem{lit6spoof}
A.~Thukral, Jyoti, and M.~Kumar, ``Iris spoofing through print attack using svm
  classification with gabor and hog features,'' in \emph{2022 International
  Conference for Advancement in Technology (ICONAT)}, 2022, pp. 1--6.

\bibitem{lit5spoof}
J.~E. Tapia, S.~Gonzalez, and C.~Busch, ``Iris liveness detection using a
  cascade of dedicated deep learning networks,'' \emph{IEEE Transactions on
  Information Forensics and Security}, vol.~17, pp. 42--52, 2022.

\bibitem{localglobalp}
C.~Yang, A.~Kortylewski, C.~Xie, Y.~Cao, and A.~Yuille, ``Patchattack: A
  black-box texture-based attack with reinforcement learning,'' in
  \emph{Computer Vision -- ECCV 2020}, A.~Vedaldi, H.~Bischof, T.~Brox, and
  J.-M. Frahm, Eds.\hskip 1em plus 0.5em minus 0.4em\relax Cham: Springer
  International Publishing, 2020, pp. 681--698.

\bibitem{gradient-free}
M.~Alzantot, Y.~Sharma, S.~Chakraborty, H.~Zhang, C.-J. Hsieh, and
  M.~Srivastava, ``Genattack: Practical black-box attacks with gradient-free
  optimization,'' 2018.

\bibitem{IrisDatset}
{ National Laboratory of Pattern Recognition (NLPR) - Institute of Automation,
  Chinese Academy of Sciences (CASIA)}, ``{CASIA Iris Subject Ageing},''
  http://biometrics.idealtest.org/, {Accessed in October 4th 2022}.

\bibitem{Github}
{Artifitialleap-MBZUAI}, ``{Incremental Training Data Attack},''
  https://github.com/Artifitialleap-MBZUAI/IncrementalTrainingDataPoisoning,
  October 2022.

\bibitem{twentyperc}
M.~Jagielski, A.~Oprea, B.~Biggio, C.~Liu, C.~Nita-Rotaru, and B.~Li,
  ``Manipulating machine learning: Poisoning attacks and countermeasures for
  regression learning,'' \emph{2018 IEEE Symposium on Security and Privacy
  (SP)}, pp. 19--35, 2018.

\end{thebibliography}
\end{document}